# An ontology-based approach for semantics ranking of the web search engines results

Abdelkrim Bouramoul*, Mohamed-Khireddine Kholladi
Computer Science Department, Misc Laboratory, University of Mentouri Constantine. B.P. 325, Constantine 25017, Algeria
a.bouramoul@yahoo.fr / Kholladi@yahoo.fr

Bich-Liên Doan
Computer Science Department, SUPELEC. Rue Joliot-Curie, 91192 Gif Sur Yvette, France
bich-lien.doan@supelec.fr

*Abstract*— **This work falls in the areas of information retrieval and semantic web, and aims to improve the evaluation of web search tools. Indeed, the huge number of information on the web as well as the growth of new inexperienced users creates new challenges for information retrieval; certainly the current search engines (such as Google, Bing and Yahoo) offer an efficient way to browse the web content. However, this type of tool does not take into account the semantic driven by the query terms and document words. This paper proposes a new semantic based approach for the evaluation of information retrieval systems; the goal is to increase the selectivity of search tools and to improve how these tools are evaluated. The test of the proposed approach for the evaluation of search engines has proved its applicability to real search tools. The results showed that semantic evaluation is a promising way to improve the performance and behavior of search engines as well as the relevance of the results that they return.**

## I. INTRODUCTION

Information Retrieval (IR) is a domain that is interested in the structure, analysis, organization, storage, search and discovery of information. The challenge is to find in the large amount of available documents; those that best fit the user needs. The operationalization of IR is performed by software tools called Information Retrieval Systems (IRS), these systems are designed to match the user needs representation with the document content representation by means of a matching function. The evaluation of IRS is to measure its performance regarding to the user needs, for this purpose evaluation methods widely adopted in IR are based on models that provide a basis for comparative evaluation of different system effectiveness by means of common resources. IR, the IRS and evaluation of IRS are three inseparable elements representing the domain where the problematic of this work is located.

In this context, several questions arise regarding the improvement of the information retrieval process, and the manner in which returned results are evaluated. So, is to find solutions for the two following questions: How can we improve information retrieval by taking semantics into account? And how can we ensure a semantic evaluation of the responses returned by information retrieval tools?

This paper is organized as follows: We present initially similar work and we give the principle of the proposed approach, we define its parameters in terms of the chosen information search model and the used linguistic resource. We present then the developed modules to build the general architecture of our proposal and we describe the developed tool. We finally present the experimental our approach and the discussion of the obtained results.

## II. RELATED WORK

### A. Ontology definition

Several definitions of the ontology have emerged in the last twenty years, but the most referenced and synthetic one is probably that given by Gruber: "ontology is an explicit specification of a conceptualization" [5]. Based on this definition, ontologies are used in the IR field to represent shared and more or less formal domain descriptions in order to add a semantic layer to the IRS.

### B. Ontologies, a clear need in IR

It is natural that works relating to ontology integration in IRS are growing. A first solution is to build ontology from the corpus on which IR tasks will be performed [8] [6]. A second solution is the reuse of existing resources. In this case, ontologies are generally chosen from the knowledge domain that they address [1], [10]. Ontologies as a support for the modeling of IRS have been studied in a previous article [2]. In general, the contribution of ontologies in an IRS can be understood at three levels:

- In the document indexing process: by combining it with the techniques of natural language processing, the documents in the database will be summarized and linked to the ontology concepts. If this step has been properly done, the search would be easier in the future. This principle was already used in our work [3].
- At the queries reformulation level in order to improve the initial user queries. This aspect was also used as a complement to our proposal [3].

- In the information filtering process, this aspect will be the subject of the contribution that we present in this paper. The idea is to use ontology to add the semantic dimension to the evaluation process. This can be done by extracting the query terms and their semantic projection using the WordNet ontology on the set of returned documents. The result of this projection is used to extract concepts related to each term, thus building a semantic vector which will be the base of the results classification. This vector is used primarily for creating the query vector and document vector used by the vectorial model that we adopted.

## III. THEORETICAL FOUNDATIONS OF THE PROPOSED APPROACH

We present in this section the theoretical basis on which our proposal is based. These features guide the semantic evaluation approach that we propose. In this paper we are interested specifically in the semantic evaluation of the results returned by search engines. For this purpose, our choice is fixed on three search engines (Google, Yahoo and Bing). This choice is motivated by their popularity in the Web community on the one hand and the degree of selectivity that they offer on the other. More precisely, our system allows to:

- Retrieve the results returned by search engines
- Check the information content of each returned page.
- Project the user query on the linguistic resource, the WordNet ontology in our case.
- Measure the results relevance by calculating the relevance degree of each of them.
- Generate a semantic rank of results according to the calculated relevance based on their degree of informativeness.
- Assign a score to each search engine based on its position in the new ranking.

This system is based partly on a linguistic resource (WordNet ontology) for the query semantic projection and on the other hand, a calculation model for measuring the relevance 'document/ query' (the vectorial model). In the following we are justifying our choices in terms of the chosen linguistic resource and the used IR model.

### A. Choice of information retrieval model

The role of an IR model is to provide a formalization of the information finding process. The definition of an information retrieval model led to the determination of a theoretical framework. On this theoretical framework the representation of information units and the formalization of the system relevance function are based.

#### 1) Summary of IR models
We have given as part of our previous work [4] an overview of the most common information retrieval models. We remind the basics of each of them in order to center our choice on the model that fits best with our proposal. Figure 1 shows the three IR model that we studied.

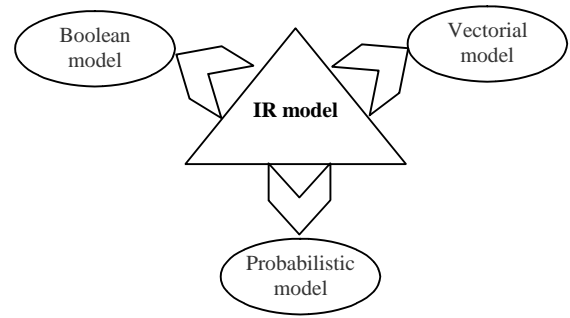

Figure 1. Information Retrieval Models

The Boolean model is based on the keywords manipulation. On the one hand a document (D) is represented by a combination of keywords, on the other hand a query (R) is represented by a logical expression composed of words connected by Boolean operators (AND, OR, NOT). The Boolean model uses the exact pairing mode; it returns only documents corresponding exactly to the query. This model is widely used for both bibliographic databases and for web search engines.

The vector model recommends the representation of user queries and documents as vectors in the space generated by all the terms. Formally, the documents and queries are vectors in a vectorial space of dimension N and represented as follows:

$$D_j = (d_{j1} d_{j2} d_{jT}) \quad Q_k = (q_{k1} q_{k2} q_{kT})$$

Finally, the probabilistic model uses a mathematical model based on the theory of probability. In general, the probabilistic model has the advantage of unifying the representations of documents and concepts. However, the model is based on assumptions of independence of variables not always verified, tainting the measures of similarity of inaccuracy.

#### 2) Principles and motivations of the chosen model
In the semantic evaluation approach that we propose, we opted for the vectorial model, this choice is mainly motivated by three reasons: first, the consistency of its representation "Query/Document", then the order induced by the similarity function that it uses, and finally the easy possibilities that it offers to adjust the weighting functions to improve search results.

More precisely in our case, the vectorial model is based on a semantic vector composed of concepts rather than words. This semantic vector is the result of the semantic projection of the query on the WordNet ontology. This model therefore allowed us to build "query vectors" and «document vectors" on the basis of coefficients calculated using a weighting function. It was also the basis for measuring the similarity between the query vector and those of documents using a calculation function of similarity between vectors. The term weighting scheme and the similarity measures used in conjunction with this model are:

*Term Weighting:* It measures the importance of a term in a document. In this context, several weighting techniques have been developed, most of them are based on "TF" and "Idf" factors [9] that combine local and global term weights:

- TF (Term Frequency): This measure is proportional to the frequency of the word in the document (local weighting).
- Idf (Inverse Document Frequency): This factor measures the importance of a term in the entire collection (total weight).

The "TF*Idf" measure gives a good approximation of the word importance in the document, especially in corpora with a similar amount of documents. However, it ignores an important aspect of the document: its length. For this reason we used the following standard formula [7]:

$$TF_{Di} = \frac{\sum occ(w)}{card\ Di} \quad (1)$$

*Similarity measure:* Two similarity measures of each document according to the same query are calculated by our system:

- The distance measure in a vectorial space:

$$Dist(Q_k, D_j) = \sum_{i=1}^{T} |q_{ki} - d_{ji}| \quad (2)$$

- The cosine measure to measure the similarity of documents and query. This measure is also called the document correlation $D_j$ relative to the query terms $Q_k$.

$$RSV(Q_k, D_j) = \frac{\sum_{i=1}^{T} q_{ki} d_{ji}}{\sqrt{\sum_{i=1}^{T} q_{ki}^2 \sum_{i=1}^{T} d_{ji}^2}} \quad (3)$$

### B. Choice of linguistic resource

We thought, initially, to use domain ontology in the medical or geographic field and exploit collections of documents related to these fields. But we realized that this kind of ontology is generally developed by companies for their own needs. At least, they are not available on the Internet. Moreover, few of them have a terminology component (terms associated with concepts). So, our choice was oriented to the WordNet ontology.

WordNet is an electronic lexical network developed since 1985 at the Princeton University by a linguists and psycholinguists team of the Cognitive Science Laboratory. The advantage of WordNet is the diversity of the information that it contains (large coverage of the English language, definition of each meaning, sets of synonyms and various semantic relations). In addition, WordNet is freely usable.

WordNet covers the majority of nouns, verbs, adjectives and adverbs of the English language. They structure it into a nodes and links network. The nodes consist of sets of synonyms (called synsets). A term can be a single word or a collocation. Table 1 provides statistics on the number of words and concepts in WordNet in its version 3.0.

TABLE I. CHARACTERISTICS OF THE WORDNET 3.0 ONTOLOGY

| Category | Words | Concepts | Total Pairs Word Sense |
|---|---|---|---|
| noun | 117 798 | 82 115 | 146 312 |
| verb | 11 529 | 13 767 | 25 047 |
| adjective | 21 479 | 18 156 | 30 002 |
| adverb | 4 481 | 3 621 | 5 580 |
| Total | 155 287 | 117 659 | 206 941 |

WordNet concepts are linked by semantic relations. The basic relationship between the terms of the same synset is the synonymy. Moreover the different synsets are linked by various semantic relations such as subsumption or hyponymy-hyperonymy relation, and the meronymy-holonymie composition relationship.

### IV. PRESENTATION OF THE PROPOSED APPROACH

In order to ensure a coherent modeling of our proposal, we have created a number of modules where each of them ensures a separate functionality. The combination of these modules has allowed us then to build the general architecture of the system. These modules are interrelated in the sense that the outputs of each module are the inputs of the next. Figure 2 shows how the different modules are connected to define the general architecture describing our approach.

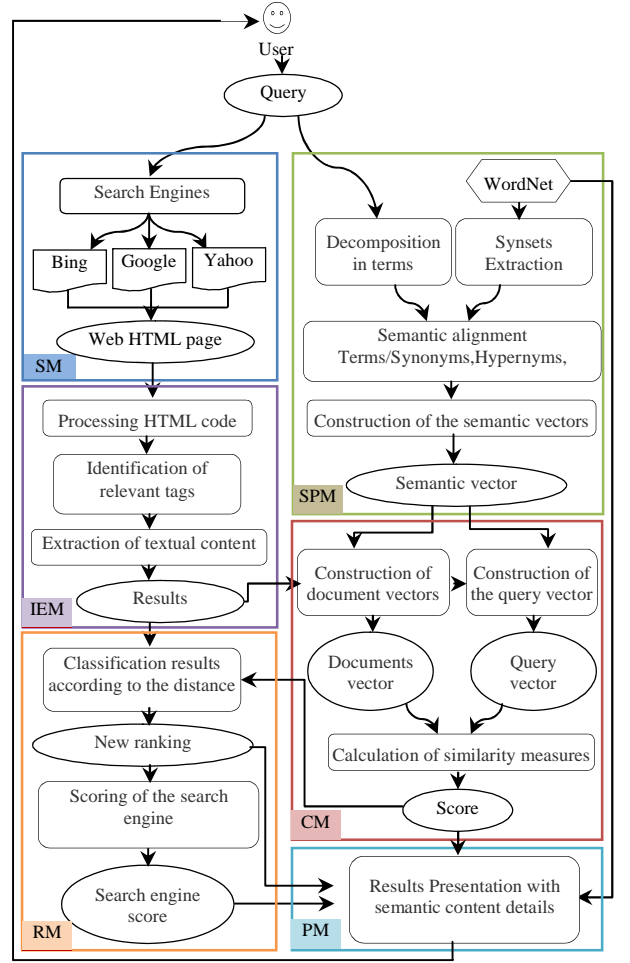

Figure 2. General architecture of the proposed approach

We will present in the following these modules, specifically we will describe the inputs, outputs and the principle of operation of each of them.

*A. Search Module (SM)*

In order to implement our proposal, our choice was fixed on the three search engines (Google, Yahoo and Bing), who now represent the most used search tools by the Web community.

The search module transmits the user query to search engines Google, Yahoo and Bing, and retrieves the first 20 responses returned by each of them. This set of results represents the information content to be evaluated. The choice of the top 20 results is justified by the fact that they represent the links that are usually visited by the user on all the returned results. They are those that contain the most relevant answers. However, we note in this context that this number can be expanded to cover all the returned results. Logically, the consequence is that the processing time will be longer in this case.

*B. Information Extraction Module (IEM)*

This module supports the extraction of information content of web pages returned by the search module. This is mainly to recover the information contained in the HTML tags describing respectively the title, abstract, and URL of each result. This treatment is performed for the first two pages containing the 20 results returned by each of the three search engines.

Indeed, the results page returned by a search engine, in its raw state, contains HTML formatting and representation tags, these latter do not provide useful information, and they should not be taken into account by the evaluation. In this context, we precede with the purification (cleaning) of the resulting html pages before collecting the URLs of pages to visit (those which are to be evaluated).

The difference in the structure and the format used by all three search engines forced us to implement an HTML parser for each of them to adapt the purification process and the recovery to the structure of the one that the engine uses. Once the purification process is complete, the page corresponding to each link is opened and its contents are treated to prepare the data for evaluation. This treatment is provided by the extraction module and includes:

– Parsing the HTML code of the current page from the URL in question.
– Treatment of HTML tags: the page code (its information content) must be processed to retrieve only the content that is behind the tags found useful in our case.

*C. Semantic Projection Module (SPM)*

In order to take semantics into account when generating the new classification, we associate with each query term the set of words that are semantically related. The idea is to project the query terms on the ontology concepts using the two semantic relations, 'synonymy' and 'hypernonymie' to extract the different query senses. Thereafter, all the concepts that are recovered for each term are used in conjunction with the term itself during the weighting by the calculation module. The aim is to promote a document that contains words that are semantically close to what the user is looking for, even if those words do not exist as terms in the query.

We use for this purpose, the WordNet ontology according to the following: Initially we access the part of the ontology containing the concepts and semantic relations, the latter are used to retrieve all synsets relating to each terms of the query. These synsets are finally used to build the semantic vector that contains for each query term the appropriate synonyms and hypernyms.

*D. Calculation Module*

Once the text content and the semantic vector are built, the calculation module performs the construction of the documents and query vectors based on coefficients calculated by using the appropriate weighting function (Formula 1). The calculation module then measures the similarity between these two vectors using the similarity calculation functions between vectors (Formula 2 and 3). The operation of this module is performed in two steps:

*a. Term weighting:* This step takes into account the weight of terms in the documents. It proceeds as follows:

– A $d_{ij}$ coefficient of the $D_j$ document vector measures the weight of term $i$ in document $j$, according to the formula (1)
– A $q_i$ coefficient of query vector $Q$ measures the weight of term $i$ in all documents.

*b. "Document/query" matching:* The comparison between the document vector and the query vector sums up to calculating a score that represents the relevance of the document regarding to the query. This value is calculated based on the distance formula (2) and the correlation formula (3).The matching function is very closely related to the query term weighting and the documents to be evaluated.

*E. Ranking Module (RM)*

The role of the similarity function is to order documents before they are returned to the user. Indeed, users are generally interested in examining the initial search result. Therefore, if the desired documents are not presented in this section of results, users consider the search engine as badly adjusted to their information needs, and the results that it returns will be considered as irrelevant. In this context, the role of the ranking module is to finalize the semantic evaluation process by adapting the system relevance to the user's one.

At this stage of the evaluation process, each document is described by two similarity values generated by the calculation module. Based on the distance between the document vector and the query vector, the ranking module performs the scheduling of the results so that the document with the lowest distance value, and therefore the higher relevance will be ranked first until all results are properly arranged.

This module also supports the relevance measure of the search engine itself. This is done by assigning to each of the three search engines (Google, Yahoo and Bing) a relevance score. This score is calculated by comparing the ranking results produced by each search engine to the new semantic ranking generated by our approach.

*F. Presentation Module (PM)*

The search engine results are generally presented as a list of links accompanied by title and abstract describing the content of each page. These results, before being presented to the user, must be ordered according to the relevance score assigned by the algorithms of each search engine.

In the approach that we propose, with respect to our general principle to display the search results, the presentation module supports the display part when the results are processed. Specifically, this module provides a summary of the search session as follows:

– All results in response to the query, where each result is represented by a triplet (title, abstract, URL). These results are semantically ranked according to the principle of the proposed approach.
– The semantic relevance score associated with each result.
– The set of concepts related to each query term. These concepts are retrieved from the WordNet ontology and presented as a tree.

## V. THE DEVELOPED TOOL

To demonstrate the applicability of the proposed approach, we have developed a tool for the semantic evaluation of the results returned by search engines. To this end, it was necessary to develop a simple interface to allow the user to perform certain checks on the current evaluation session. This interface is based on the following components:

– The global view that summarizes the state and the initial ranking of all the responses returned by the three search engines Google, Yahoo and Bing.
– The formulation of the query and the various concepts after its projection on ontology.
– The ability to choose the type of ranking to be made.

Figure 3 shows the main window of this tool.

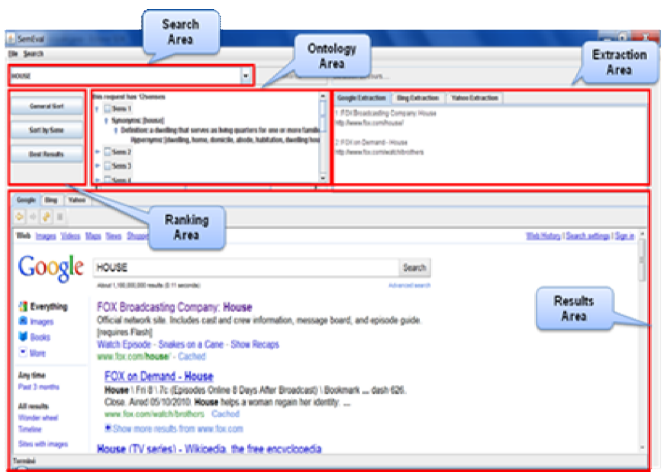

Figure 3. The developed tool

## VI. TEST OF THE PROPOSED APPROACH

*A. The used method*

The objective of this experimentation is to measure the contribution of the inclusion of semantics in the ranking of results returned by search engines. The idea is to display results according to two different ways: first, a default ranking as was proposed by the search engine we call 'classical ranking' and a second ranking generated by our system scheduling results according to the ontology-driven approach that we propose, we refer to this ranking by 'semantic ranking'. This test aims to measure the users' satisfaction by comparing for the same set of queries both types of result rankings.

To this end, we are interested in the first 20 results to measure each search engine performance according to the two ranking types (classical and semantic). We also treated the case of redundant results, parasite links and dead links. We have studied the results of Google and Yahoo from a series of 25 search scenarios including 15 simple scenarios covering the range of current needs of a user (they were simple applications of thematic travel, consumption, news and culture) and 10 complex scenarios (rare word or specialized search). In total 25 queries and 500 results were screened to a scoring grid.

*B. Results and Discussion*

*1) General Performance*

TABLE II. THE EFFECTIVENESS COMPARISON OF TWO SEARCH ENGINES

|  | Google | | Yahoo | |
| --- | --- | --- | --- | --- |
|  | **Classical** | **Semantic** | **Classical** | **Semantic** |
| **Overall average** | 7,62 | 8,29 | 6,93 | 7,02 |
| **Simple scenarios** | 8,15 | 8,82 | 7,76 | 7,89 |
| **Complexes scenarios** | 6,19 | 6.94 | 5,23 | 5,52 |

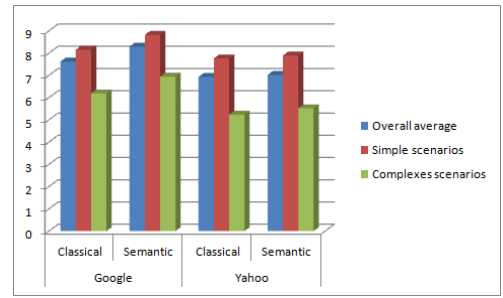

Figure 4. The effectiveness comparison of two search engines

This first result confirms the quality of Google which returns the best services to the user; Google had scored higher on almost all the queries made. But the difference of the overall average to Yahoo is not significant: only 0.69 of 10 points in the case of classical ranking and 1.27 for the semantic ranking separate the two search engines. And this difference is reduced to 0.43 and 0.93 point in the case of simple queries whereas it increases in the case of complex search scenarios (0.96 and 1.42 point). We also find that the three criteria and in the case of the two search engines, semantic ranking always brings a gain in efficiency compared to the classical one.

*2) Performance by criteria*

TABLE III. COMPARISON OF THE TWO SEARCH ENGINES EFFECTIVENESS BY CRITERIA

|  | Google | | Yahoo | |
| --- | --- | --- | --- | --- |
|  | Classical | Semantic | Classical | Semantic |
| The results relevance | 5,72 | 6,12 | 5,06 | 7,66 |
| Not dead links | 9,60 | 9,67 | 9,11 | 9,32 |
| Non-redundant results | 8,27 | 7,92 | 7,55 | 7,02 |
| Not parasites pages | 9,33 | 9,37 | 8,59 | 8,86 |

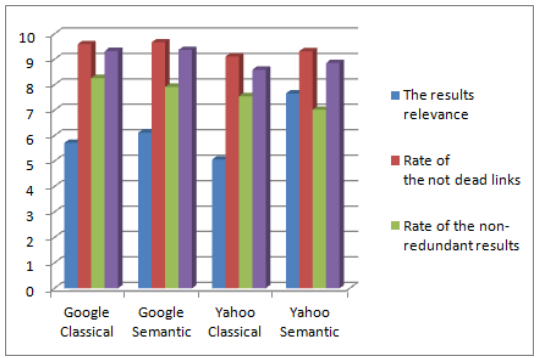

Figure 5. Comparison of the two search engines effectiveness by criteria

With respect to the results relevance, the difference between the two search engines (0.66 point for classical ranking and 1.54 for semantic ranking) is remarkably larger than that of the total score. This is explained in particular by the more relevant results for complex searches in Google. However, both are above the average for that criterion. We also note that for both search engines, the semantic ranking improves the relevance of the results especially in the case of Yahoo, where the gain in terms of relevance amounts to 2.60 points.

Regarding to the dead links level, the test reveals the effort of the two engines to maintain their index and avoid pointing to deleted or moved pages. On this criterion very clearly Google precedes Yahoo for 0.49 and 0.34 point. This criterion shows a slight advance of the semantic ranking compared to the classical one.

In terms of redundant results, Google and Yahoo are doing well. Ergonomically, Google gets a higher score with a more relevant outcome: When it displays on a page two links that point to the same site (but different pages), it takes care to paste the two results and displays the second with a slight shift to the right. Visually, the user can see that the two results are related. Contrary to what was expected for this criterion, the classical ranking gives better scores compared to the semantic one; because the number of synonyms retrieved from the ontology increases the frequency of query terms in the returned documents, which promotes links arriving from the same site.

Regarding to the parasite pages (pages listing only promotional links), Google is more effective than Yahoo to deal this kind of useless pages in advancing the user search otherwise these distort engine results (as merely advertising and often poorly targeted) . Scores are 9.33 and 8.59 for the classical ranking and 9.37 and 8.86 for the semantic one, so we see a better result in the case of semantic ranking.

## VII. CONCLUSION

In this paper, we have presented our contribution for the semantic evaluation of results returned by search engines. This approach is not specific to a particular type of research tool; it is rather generic because the ontology that we used is not specific to a particular domain.

The structuring of the proposed approach into a set of modules aims to define a modular and rchitecture in the sense that any adjustment or change in one module does not affect the functioning of other modules. Our proposal consists of six modules that provide the following functionality: First, the recovery of web pages containing the responses of search engines and the extraction of information that will be evaluated. Thereafter it will project the query terms on the concepts of the ontology. The evaluation itself has to construct documents and query vectors to generate a semantic ranking of results returned by search engines according to the used similarity functions. Finally, the results of the evaluation are presented to the user.